\documentclass[fleqn,10pt]{wlscirep}
\usepackage{amsmath}
\usepackage{amssymb}
\usepackage{gensymb}
\usepackage{xargs}
\usepackage{dsfont}
\usepackage{color}
\usepackage{amsfonts}
\usepackage{bm}
\usepackage{graphicx}
\usepackage{upgreek}
\usepackage{amsmath,amssymb}
\usepackage{hyperref}
\usepackage{dcolumn}
\usepackage{bm}
\usepackage{xcolor}
\usepackage[colorinlistoftodos]{todonotes}

\usepackage{soul}

\usepackage{multirow}  
\usepackage[left]{lineno}

\begin{document}
\title{Batch Bayesian optimization of attosecond betatron pulses from laser wakefield acceleration}

\author[1*]{Dominika~Maslarova}
\author[1]{Albert~Hansson}
\author[1,2]{Mufei~Luo}
\author[1,3]{Vojtěch~Horn\'y}
\author[1]{Julien~Ferri}
\author[1]{Istvan~Pusztai}
\author[1]{T\"unde~F\"ul\"op}
 
\affil[1]{Department of Physics, Chalmers University of Technology,  G\"{o}teborg, SE-41296, Sweden}%

\affil[2]{Department of Physics, University of Oxford, Parks Road, Oxford OX1 3PU, UK}%
\affil[3]{Extreme Light Infrastructure - Nuclear Physics, Horia Hulubei National Institute for Physics and Nuclear Engineering, 30 Reactorului Street, RO-077125 Bucharest-Magurele, Romania }%
\affil[*]{E-mail: dommas@chalmers.se}%

\begin{abstract}    
Laser wakefield acceleration can generate a femtosecond-scale broadband X-ray betatron radiation pulse from electrons accelerated by an intense laser pulse in a plasma. The micrometer-scale of the source makes wakefield betatron radiation well-suited for advanced imaging techniques, including diffraction and phase-contrast imaging. Recent progress in laser technology can expand these capabilities into the attosecond regime, where the practical applications would significantly benefit from the increased energy contained within the pulse. Here we use numerical simulations combined with batch Bayesian optimization to enhance the radiation produced by an attosecond betatron source. The method enables an efficient exploration of a multi-parameter space and identifies a regime in which a plasma density spike triggers the generation of a high-charge electron beam. This results in an improvement of more than one order of magnitude in the on-axis time-averaged power within the central time containing half of the radiated energy, compared to the reference case without the density spike.

\end{abstract}
\maketitle
\section*{Introduction}
The generation of electromagnetic pulses on attosecond timescales represents a milestone in ultrafast science. These pulses offer unprecedented time resolution, enabling real-time observation of electron dynamics in atoms and molecules. A major advancement in attosecond pulse generation has been achieved through high-order harmonic generation~\cite{mcpherson1987studies,ferray1988multiple, l1991higher, lewenstein1994theory}, where intense laser fields produce extreme ultraviolet and soft X-ray radiation. Attosecond pulses can also be generated with femtosecond-scale intense laser pulses through electron acceleration and associated betatron radiation, as demonstrated by numerical simulations~\cite{ horny2020attosecond, ferri2021generation}. Betatron radiation is produced during laser wakefield acceleration (LWFA) \cite{tajima1979laser}, a laser-plasma interaction technique for accelerating electrons to relativistic speeds.
 While LWFA electron beams typically have durations on femtosecond timescales - corresponding to the typical dimensions of the wakefield structure - it is also possible to push the beam durations into the sub-fs regime \cite{luttikhof2010generating, tooley2017towards, zhao2019sub,horny2020attosecond,ferri2021generation,kim2021subfemtosecond, deng2023generation, tomassini2023attosecond,sun2024generation, tomassini2025ultra}, with betatron radiation inheriting the attosecond scale~\cite{horny2020attosecond,ferri2021generation}. 
 
 LWFA betatron radiation, characterized by broadband X-ray spectra \cite{albert2016applications,nemeth2008laser,cipiccia2011gamma}, offers good spatial coherence, as demonstrated in femtosecond-scale X-ray phase contrast imaging of biological \cite{kneip2011,fourmaux2020laser,cole2015laser} and complex microstructure \cite{hussein2019laser} samples. Extending photon energies of attosecond betatron pulses into the keV range enables X-ray absorption spectroscopy of high-Z elements \cite{seres2006x}, and simultaneous multi-element probing in single-shot measurements \cite{yamada2021broadband}. Attosecond betatron radiation thus has the potential to significantly complement techniques like high-harmonic generation to enhance temporal resolution and spatial coherence, improving the ability to focus X-ray beams for high-resolution imaging and precise measurements in ultrafast science.

In LWFA, an ultrashort, ultraintense laser pulse travels through an underdense plasma medium, driving a plasma wave (wakefield) that creates an accelerating gradient that is of the order of a thousand times higher than that of conventional radiofrequency accelerators. The wakefield wavelength is on the order of the plasma wavelength, $\lambda_{\mathrm{p}}=2 \pi c  \sqrt{\epsilon_0 m_e/(e^2 n)}$, where $c$ is the speed of light, $\epsilon_0$ is the vacuum permittivity, $m_e$ is the electron mass, $e$ is the elementary charge and $n$ is the electron density. For laser intensities corresponding to normalised vector potential $a_0=0.855\lambda_{\mathrm{L}} [\mathrm{\upmu m}] \sqrt{I_{\mathrm{L}} [10^{18} \mathrm{W~cm^{-2}}}]\gtrsim 2$, where $I_{\mathrm{L}}$ is the laser intensity and $\lambda_{\mathrm{L}}$ is the laser wavelength, the radiation pressure of the laser pulse expels plasma electrons radially outward, creating ion cavities surrounded by electron sheaths. The first cavity following the laser pulse, the ``bubble'', offers the strongest acceleration. This highly nonlinear regime of acceleration is therefore also referred to as the bubble regime.
The accelerating field is located at the rear half of the bubble with respect to the pulse propagation direction. During the acceleration in the bubble, the electron beam undergoes transverse betatron oscillations due to the presence of transverse focusing forces, making the bubble also serve as a wiggler. This mimics the principle of synchrotron devices\cite{kiselev2004x, rousse2004production}, where the radiated power generally increases with particle energy and the curvature of the trajectory. 
{In the case of the betatron radiation in the bubble regime \cite{kiselev2004x, rousse2004production}, both the number of photons and the critical energy -- below which half of the radiation power is emitted -- increase with the charge of the electron beam, electron energy and oscillation amplitude \cite{corde2013femtosecond}.}
 
Practical applications generally benefit from increased energy contained within the pulse, as this leads to a stronger detectable signal in imaging.
Several methods to improve the properties of the betatron radiation have been demonstrated with femtosecond electron beams \cite{nemeth2008laser,cipiccia2011gamma,huang2016resonantly,mangles2006laser,chen2013bright, mangles2009controlling, popp2010all, wood2017enhanced, yan2014concurrence, zhang2016enhanced, ho2013, wallin2017radiation, shaw2014role,zhao2016high,ferri2018enhancement,ma2018angular,ta2008betatron, guo2019enhancement, ferri2018high}.
For instance, tailoring the plasma density through transverse density gradients \cite{ferri2018enhancement,ma2018angular}, longitudinal density shaping \cite{ta2008betatron, guo2019enhancement}, or additional high-density stages \cite{ferri2018high} can enhance betatron emission. A density increase along the propagation direction shortens the plasma wavelength and contracts the bubble, shifting electrons toward the rear of the cavity where the accelerating fields intensify, enabling a phase-reset that renews their energy gain \cite{dopp2016energy} and affects the properties of the betatron radiation \cite{ta2008betatron}.

It is generally challenging to optimize several laser-plasma parameters that can affect the properties of betatron radiation at once due to the high computational cost of kinetic simulations. Bayesian optimization offers an efficient solution by using a probabilistic surrogate model to capture the relationship between adjustable input parameters of the laser-plasma system, and output parameters quantifying the properties of the  electron beam and the betatron radiation. In LWFA, it has been applied to improve electron beams in experiments \cite{shalloo2020automation,jalas2021bayesian, jalas2023, irshad2024pareto} and numerical simulations \cite{jalas2021bayesian, ferran2023bayesian,jalas2023, irshad2023,zhong2024, valenta2025bayesianoptimizationelectronenergy}, as well as to enhance betatron-source performance \cite{shalloo2020automation,ye2022fast}. 
These studies have shown that optimizing parameters such as laser focal position and plasma density \cite{shalloo2020automation,ye2022fast}, laser spectral phase and plasma length \cite{shalloo2020automation} can substantially increase betatron photon yield and improve X-ray imaging quality.

In this study, we employ batch Bayesian optimization (BBO) to improve the betatron source on an attosecond scale, using three-dimensional (3D) particle-in-cell (PIC) simulations. BBO is an extension of Bayesian optimization that simultaneously selects multiple input points in one iteration, as opposed to evaluating one point at a time, which is practical for a parallelization of the data acquisition when the simulations are time consuming. Using a density spike in the plasma structure and optimizing its position and gradients, we observe more than $25$ times enhancement of the peak of the on-axis radiated energy and more than $6$ times increase of the energy contained within the central 50\% of the betatron pulse, with only a few tens of PIC simulations requiring a few thousand core hours each. The simulations demonstrate concentrated betatron radiation generation on a millijoule-class laser system using a plasma target, highlighting the potential of low-cost, compact setups for producing attosecond X-ray pulses.

  \section*{Results}\label{sec:optimization}
  \subsection*{Enhancement of the betatron radiation} In our simulations, the attosecond electron beam is produced and injected into the wakefield by a down-ramp injection method \cite{bulanov1998particle}, where a density gradient is applied at the plasma entrance. The gradient changes the wake phase-velocity, leading to wave breaking, where electrons outrun the rear part of the bubble and consequently are injected into it, where they can be further accelerated. 
  The betatron enhancement {was expected to occur} due to a locally increased plasma density after a distance $d_u$, which shifts electrons to the rear part of the bubble, allowing them to gain more energy during the ``phase-reset'' mechanism \cite{dopp2016energy}. 
  
\begin{figure}[htbp]
    \centering
   \includegraphics[width=0.5\textwidth]{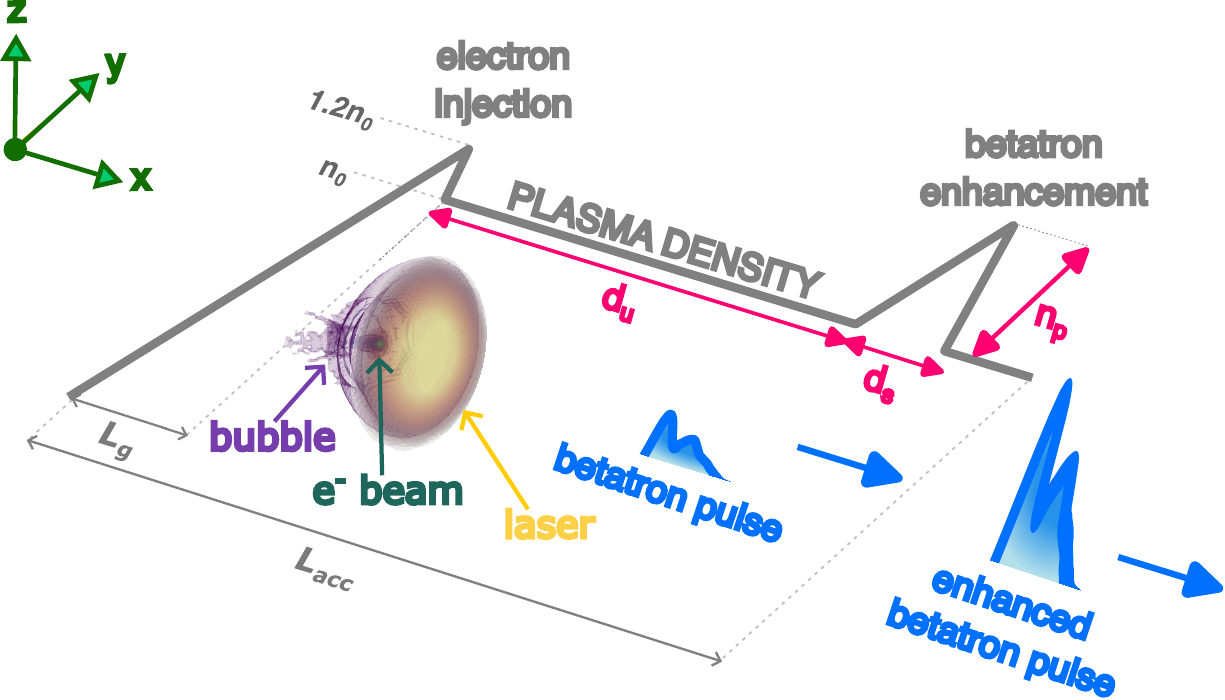}
    \caption{\textbf{An illustration of the proposed setup for betatron radiation enhancement.} The plasma density profile (gray thick line) varies along the longitudinal coordinate $x$, while remaining constant in the $y$ and $z$ coordinates. The laser pulse first enters a gradient of length $L_\mathrm{g}=40~{\mathrm{\upmu m}}$, providing the injection of the electron ($e^-$) beam into the bubble. The $e^-$ beam produces a betatron pulse that is enhanced by a density spike introduced further along $x$. In the simulations, the length of the density spike $d_{\mathrm{s}}$, the length of the uniform region with density $n_0$ before the spike $d_{\mathrm{u}}$ and the maximum density value of the density spike $n_{\mathrm{p}}$ were adjusted in the optimization process. The total plasma length is equal to $L_\mathrm{acc}=256~{\mathrm{\upmu m}}.$}\label{fig:setup}
\end{figure}

  We optimize the distance between the injection of the electrons and the start of this density spike $d_{\mathrm{u}}$, the length of the spike $d_\mathrm{s}$, and its peak density $n_\mathrm{p}$, as shown in Fig.~\ref{fig:setup}. The optimization results indicate that the central value of the betatron energy increases significantly, with a simultaneous reduction in pulse duration when the spike peak density is four times higher than the reference uniform density, the density spike is located shortly (few $\mathrm{\upmu m}$) after the injection gradient at the entrance and the spike length is about $120~\mathrm{\upmu m}$. The enhancement is actually driven by a second intense electron injection, which produces the significantly more pronounced radiation peak compared to the reference, no-spike configuration.

\subsection*{Attosecond pulse generation}  
We consider the interaction of a laser pulse, propagating in the $x$-direction, with an underdense plasma. The laser has a wavelength $\lambda_{\mathrm{L}}=800~\mathrm{nm}$ and peak intensity $2.53\times10^{19}~\mathrm{W~cm^{-2}}$, corresponding to a normalised vector potential $a_0=3.44$ and pulse energy $E_{\mathrm{L}}=37~\mathrm{mJ}$. It has a Gaussian spatial and temporal profile, with linear polarization along the $y$-axis. The pulse is focused at the plasma entrance to a spot size of 3.8~$\mathrm{\upmu m}$ and has a duration of 8.3 fs (full width at half maximum (FWHM) of the intensity). The set-up is illustrated in Fig.~\ref{fig:setup}. At the entrance of the plasma, a region with longitudinal density variation of total length $L_\mathrm{g}=L_\mathrm{g_1}+L_\mathrm{g_2}=40~\mathrm{\upmu m}$  is introduced to trigger the initial electron injection. The plasma density rises linearly from zero to $ 1.2 n_0$, with $n_0=2\times10^{19}~\mathrm{cm^{-3}}$, over the first $L_\mathrm{g_1}=30~\mathrm{\upmu m}$, and then decreases back to $n_0$ over the following $L_\mathrm{g_2}=10~\mathrm{{\upmu}m}$. This configuration was used also by 
 Ferri et al.~\cite{ferri2021generation}, to  numerically demonstrate the generation of an attosecond betatron pulse from a down-ramp injection in LWFA.

To enhance the radiation output, we introduce an additional density spike starting at the position $x=L_\mathrm{g}+d_\mathrm{u}$. 
For simplicity,  the linear up and down ramps of the spike are chosen to be symmetric, 
with a density reaching its maximum value $n=n_\mathrm{p}$ in the middle of the $d_\mathrm{s}$ long region. 
In the following, the set of values \{$d_{\mathrm{u}}$, $d_{\mathrm{s}}$, $n_{\mathrm{p}}$\} varied during the optimization are referred to as a point in the parameter space. We set the total plasma length in the $x$ direction to the dephasing length $L_{\mathrm{d}} = 2{\lambda_{\mathrm{p}}}_0^3\sqrt{a_0}/(3 \pi \lambda_{\mathrm{L}}^2)=256~\mathrm{\upmu m}$, which is the distance at which the electrons outrun the accelerating wakefield region \cite{lu2007generating}. Here, ${\lambda_{\mathrm{p}}}_0$ is the plasma wavelength at the density $n_0$. Despite non-uniform density in the simulations, $L_{\mathrm{d}}$ reliably estimates the acceleration length $L_{\mathrm{acc}}$.

To simulate the production of attosecond pulses, we employ a combination of two codes. The interaction of the laser pulse and the plasma is modeled using 3D PIC simulations with Smilei\cite{derouillat2018smilei}. Based on the electron trajectories obtained from Smilei, the betatron radiation is computed using the Far-field Intensity through Kinetic Analysis (FIKA) code \cite{FIKA}, a Python-based computational tool we developed to analyze the far-field radiation profile from accelerated particles. 

\subsection*{Optimization of betatron radiation}
The goal of the optimization is to maximize the radiated energy of the attosecond betatron pulse, which is primarily directed along the pulse propagation direction, i.e., the $x$ axis. For this reason, we aim to increase the value of $\frac{\mathrm{d}^2W}{\mathrm{d}t \mathrm{d}\mathrm{\Omega}}\big|_{\text{on-axis}}$, which is the on-axis radiated energy $W$ per time $t$ per solid angle $\Omega$. At the same time, we must ensure that the duration of the betatron pulse remains in the attosecond range, similarly to our reference case that has no density spike. To capture this trade-off, the cost function, aimed to be minimized, is therefore designed as follows:
\begin{equation}\label{eq:cost}
C(W_{50},\tau_{50})=-\frac{W_{50}}{\tau_{50}}, 
\end{equation}
where $W_{50}$ is the energy per solid angle contained within the central 50\% of $\frac{\mathrm{d}^2W}{\mathrm{d}t \mathrm{d}\mathrm{\Omega}}\big|_{\text{on-axis}}$, i.e., between the 25\% and 75\%  of the cumulative curve, and $\tau_{50}$ is the time interval over which $W_{50}$ is emitted. The “betatron pulse” is defined as the portion of the signal that ends once the $\frac{\mathrm{d}^2W}{\mathrm{d}t \mathrm{d}\mathrm{\Omega}}\big|_{\text{on-axis}}$, after reaching its peak, stays below 1\% of that peak for at least 5 as. With this choice of the cost function, longer pulse durations are penalized. Constructing the cost function as a ratio is advantageous, as then a (potentially arbitrary) normalization of the individual figures of merit is not required. Another advantage is that the cost function, due to its integral form, remains stable even when the raw radiation signal contains small oscillations, and therefore does not require any additional smoothing of the radiation profile.

Note that in the reference case additional, weaker betatron pulses appear at later observation times. In our simulation setup we observe two such secondary peaks. To verify their physical origin and relative contribution, we performed an extended reference simulation covering eight wakefield periods, in which betatron radiation is generated during each plasma period. In the later periods, however, the accelerating fields are significantly weaker, resulting in electron beams with lower charge and energy. Consequently, the combined radiation from the extra seven periods amounts to only around 14\% of the betatron on-axis energy. The cost function in Eq.~\eqref{eq:cost} is defined to focus on improving the main first, strongest betatron pulse, which is the pulse of primary interest in this study. As a consequence, the secondary pulses are neither rewarded nor penalized. Optimizing these later pulses would require a different metric depending on a specific application aim, and is beyond the scope of the present work.

For each iteration of the BBO, we performed a simulation batch consisting of a set of $N$ Smilei electron-trajectory simulations followed by FIKA radiation calculations. One batch corresponds to a complete Smilei–FIKA pipeline (see Methods section and Supplementary Note 1 for more details on the optimization process). Unlike the sequential approach, where the cost function model is updated after each individual evaluation, BBO updates the model only after a full batch has been evaluated. This allowed us to run multiple simulations in parallel and simultaneously obtain the values of $W_{50}$ and $\tau_{50}$ from all of them. After each iteration, the BBO model for each $N$ was updated manually with the new $W_{50}$ and $\tau_{50}$ values, and a new batch of simulations was run. This update step required only a few minutes and the time was comparable for each $N$. 

BBO provided the flexibility to manually inspect the calculations after the evaluation, while avoiding the long wait times typically associated with updating the model after every evaluation. This approach also enabled close monitoring of the optimization process. This is particularly relevant for strongly nonlinear processes, where it is difficult to a priori know that limiting the search space in a specific way will not lead to solutions of significantly different character -- posing different resolution requirements, being physically uninteresting or poorly quantified by the cost function.

To limit the search space to relevant regions, we imposed the following parameter restrictions:
\begin{align} 
    d_\mathrm{u} &\in \left[ 0, L_{\mathrm{acc}}-L_\mathrm{g} - {d_\mathrm{s}}^{\mathrm{min}}\right],   \label{eq:con1} \\
    d_\mathrm{s} &\in \left[ {d_\mathrm{s}}^{\mathrm{min}}, L_{\mathrm{acc}} - L_\mathrm{g} \right],  \label{eq:con2}  \\
    n_\mathrm{p} &\in \left(n_0, 4n_0 \right], \label{eq:con3} \\
    d_\mathrm{u} + d_\mathrm{s} &\leq L_{\mathrm{acc}} - L_\mathrm{g}. \label{eq:con4}
\end{align} 
Here, we set the minimum spike length to ${d_\mathrm{s}}^{\mathrm{min}}=5~\mathrm{\upmu m}$.
This small value was arbitrarily chosen to avoid exploring the density profile with $d_{\mathrm{s}}=0$, which corresponds to the reference case. For $n_p$, we selected densities greater than $n_0$ with an upper threshold of $4n_0$. This value corresponds to the density where the wakefield wavelength shrinks to ${\lambda_\mathrm{p}}_0/2$, matching the acceleration length in the bubble (the bubble radius) in the plasma density of $n_0$. The condition in Eq.~\eqref{eq:con4} ensures that the combined length of the constant density and spike regions does not exceed the total acceleration length. This guarantees that the entire density spike fits within the acceleration length. Because the search space allows for many reasonable choices of parameter bounds, we fixed the bounds a priori and did not optimize them in advance. The results therefore reflect improvements within this specified domain. Alternative or expanded limits could potentially lead to other enhancement schemes - for example, higher plasma density for the accelerator-radiator arrangement \cite{ferri2018high} or longer plasma favoring additional interaction of the electrons with the laser pulse \cite{nemeth2008laser,cipiccia2011gamma,huang2016resonantly,mangles2006laser}. While such alternative configurations might yield additional improvements, analyzing these dependences lies beyond the present scope of this work.

To initiate the optimization process, we first simulated the reference case without the spike, i.e., with $d_s=0$, and obtained the reference values of $W_{50}$ and $\tau_{50}$: $W_{50}^{\mathrm{ref}}=49.3\mathrm{~nJ~sr^{-1}}$ and $\tau_{50}^{\mathrm{ref}}=437~\mathrm{as}$, respectively. From these values, we calculated the reference cost from Eq. \eqref{eq:cost} as $C^{\mathrm{ref}}=-112.8\mathrm{~MW~sr^{-1}}$. Subsequently, in the zeroth iteration, we selected $N_0=8$ initial points using a Sobol pseudo-random generator. Sobol sampling provides more uniform coverage of the three-dimensional search space than random sampling, avoiding clustering and improving exploration in the early stages. We then performed eight optimization iterations using batch sizes of $N=1,~4$ and $8$ (for a total of $N_0+N\times 8$ simulations) in order to compare their convergence and performance. The case $N = 1$ corresponds to the conventional sequential Bayesian optimization scheme. 

\begin{figure}[h]
    \centering
    \hspace{-0.8cm}
    \includegraphics[width=0.45\textwidth]{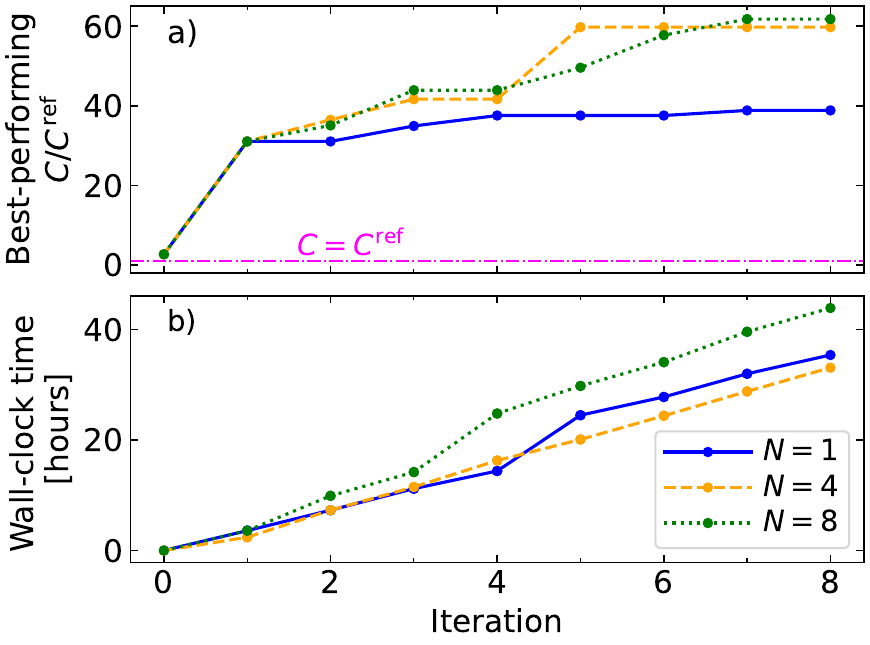}
    \hspace{-0.8cm}
    \caption{\textbf{Convergence and wall-clock time of BBO for different batch sizes.}
a) Evolution of the best-performing normalised cost value $C/C^{\mathrm{ref}}$ achieved up to corresponding iteration for different batch sizes: $N=1$ (solid blue), $N=4$ (dashed yellow), $N=8$ (dotted green). The dash-dotted magenta line marks the reference baseline, i.e., the case with $C/C^{\mathrm{ref}} = 1$.
    b) Accumulated computational wall-clock time versus iteration for different batch sizes.}\label{fig:improvement}
\end{figure}

The cost function $C$ corresponding to the best-so-far performing case for each iteration (i.e. the lowest value of Eq.~\eqref{eq:cost} found up to the corresponding iteration) are shown in  Fig.~\ref{fig:improvement}a, for the batch sizes of $N=1,~4$ and $8$. Among the tested batch sizes, $N = 4$ shows the fastest improvement within the eight performed iterations. Its best value, $C_{N=4}^{\mathrm{opt}}=60.0~C^{\mathrm{ref}}$ was reached already in the 5th iteration. After that, no further improvement was observed. For $N=8$, the best-performing case appeared later, in the 7th iteration, and was only slightly higher than for $N=4$: $C_{N=8}^{\mathrm{opt}}=62.1~C^{\mathrm{ref}}$.

In the first iteration, all batch sizes identified the same best-performing point, providing an improvement of more than $30\times$ compared to the reference. Since only one simulation was required, $N=1$ was the most efficient at this stage in terms of computational resources. However, after the next seven iterations, the $N = 1$ case reached values comparable to the second-iteration results of $N = 4$ and $N = 8$, which means that for similar number of simulations, all batch-sizes gave similar results.

Figure \ref{fig:improvement}b shows the wall-clock time measured for each batch, where the time corresponds to the interval between the start of the first simulation in the batch and the completion of the longest-running one. Overall, the total times are comparable because the simulations were run in parallel. The average wall-clock time per iteration is 4.4, 4.1 and 5.5 hours for $N=1,4$ and $8$, respectively.

Cases with better-performing cost values tend to run longer, as they typically involve processing more electrons, as discussed later. In general, however, most variations in wall-clock time across iterations and batch sizes were driven by fluctuating queue delays on the supercomputing system rather than by differences among the simulations themselves. This shows that the performance of BBO depends on available parallelization. 
Note that increasing the batch size leads to higher total CPU usage, since more simulations run simultaneously.

\begin{figure}[h]
    \centering
    \hspace{-0.8cm}
    \includegraphics[width=0.99\textwidth]{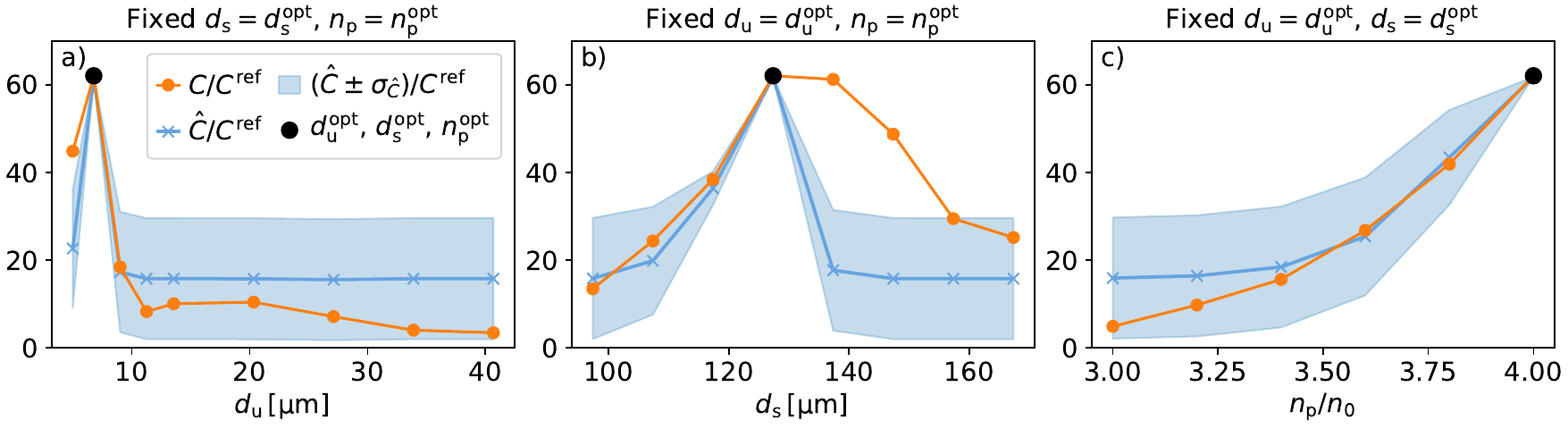}
    \hspace{-0.8cm}
    \caption{ \textbf{Values of the normalised cost function around the best-performing point found by BBO.} The values are shown for a) varying $d_{\mathrm{u}}$, fixed $d_\mathrm{s}=d_\mathrm{s}^\mathrm{opt}$, $n_\mathrm{p}=n_\mathrm{p}^\mathrm{opt}$, b) varying $d_{\mathrm{s}}$, fixed $d_\mathrm{u}=d_\mathrm{u}^\mathrm{opt}$, $n_\mathrm{p}=n_\mathrm{p}^\mathrm{opt}$, and c) varying $n_\mathrm{p}/n_0$, fixed $d_\mathrm{u}=d_\mathrm{u}^\mathrm{opt}$, $d_\mathrm{s}=d_\mathrm{s}^\mathrm{opt}$. The normalised cost function $C/C^{\mathrm{ref}}$ is shown with orange lines and circular markers at the evaluated points. The normalised model prediction $\hat{C}/C^{\mathrm{ref}}$ for $N=8$ from the final (8th) iteration, is shown with blue lines with crosses at evaluated points. Colored shaded regions indicate the area within predicted standard deviation $\sigma_{\hat{C}}$. The large black circle marks the best-performing case.}\label{fig:parameter_scan}
\end{figure}

Across all batch sizes $N=1,4$ and $8$, the best-performing parameters varied only slightly, $d_{\mathrm{u}}=6.8-8.4~\mathrm{\upmu m}$, $d_{\mathrm{s}}=112-127~\mathrm{\upmu m}$, while $n_{\mathrm{p}}$ remained consistently at $4n_0$. We performed a local parameter scan around the best-performing optimized point across all batch sizes, for which $d_\mathrm{u}=d_\mathrm{u}^{\mathrm{opt}}=6.8~\mathrm{\upmu m}$ and $d_\mathrm{s}=d_\mathrm{s}^{\mathrm{opt}}=127.4~\mathrm{\upmu m}$ and $n_\mathrm{p}=n_\mathrm{p}^{\mathrm{opt}}=4.0~n_0$, found in the 7th iteration of a batch search with $N = 8$. The scan, depicted in Fig.~\ref{fig:parameter_scan} shows that $d_{\mathrm{u}}$ is a highly sensitive parameter. Once it exceeds roughly the bubble-scale length at $n_0$ ($\lambda_{\mathrm{p}}=7.5~\mathrm{\upmu m}$), the cost drops rapidly below  $20~C^{\mathrm{ref}}$ (Fig.~\ref{fig:parameter_scan}a), which shows that the spike must follow the initial gradient within approximately one bubble scale length. In contrast, $d_{\mathrm{s}}$ and $n_{\mathrm{p}}$ (Figs. \ref{fig:parameter_scan}b and c) show a much more gradual decrease and remain robust across a wider range. Aside from slightly underestimating the performance at larger $d_{\mathrm{s}}$, overall, the final BBO model agrees well with the evaluated data. Note that, as shown in Supplementary Note 1, Fig. 1, the uncertainty remains large outside this vicinity, around $\sigma_{\hat{C}}\approx 14$. Therefore, it is generally possible that additional enhancement regions could have been found outside the best-performing region identified by our model if the exploration had been boosted by slightly modifying the optimization strategy. Exploring such possibilities lies beyond the scope of the present work.

\begin{figure*}[htbp]
    \centering
        \includegraphics[width=0.98\textwidth]{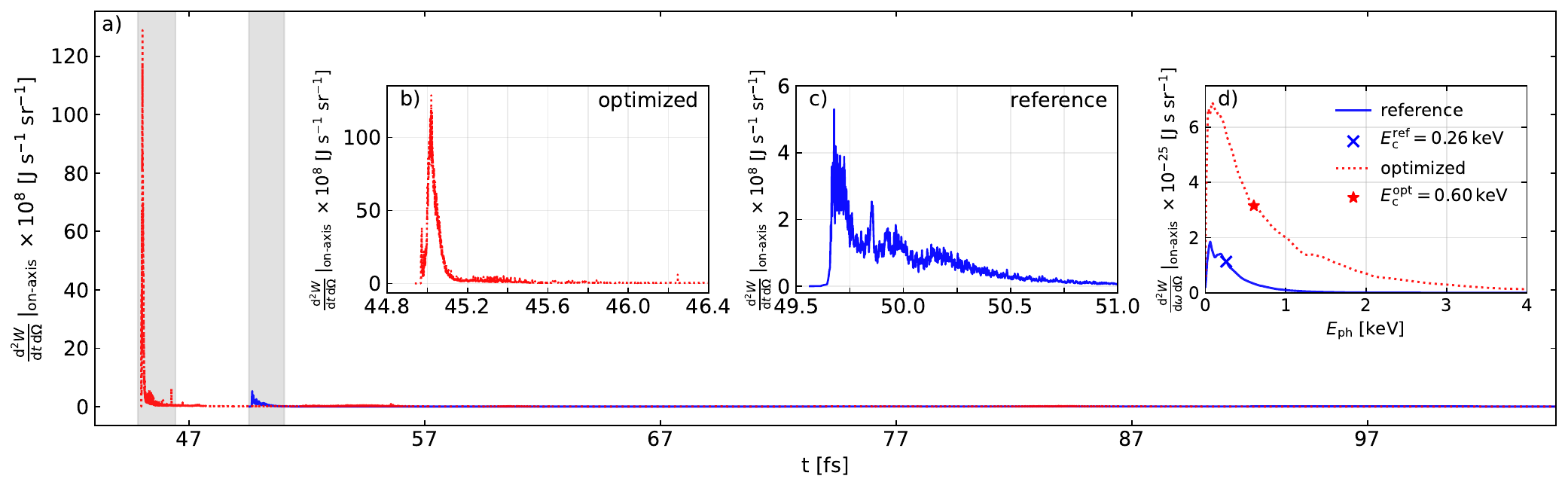}
   \caption{\textbf{Betatron radiation characteristics for the reference and best-performing cases.} a) On-axis radiated energy $W$ per time $t$ per solid angle $\Omega$ (temporal profile) as a function of the observer's time $t$ for the best-performing optimized (red dotted) and reference (blue solid) cases. The first two insets show zoomed-in profiles of the betatron peaks in b) the best-performing and c) reference cases, corresponding to the shaded areas in a). d) On-axis radiated energy $W$ per photon frequency $\omega$ per solid angle $\Omega$ of the betatron pulses as a function of the photon energy $E_{\mathrm{ph}}$. Critical energies for the reference case $E_c^{\mathrm{ref}}$ and the best-performing case $E_c^{\mathrm{opt}}$ are marked with a blue cross and a red star, respectively. The energy spectra are calculated from the whole temporal profile depicted in a). }
\label{fig:parameters_betatron} 
\end{figure*}

The characteristics of the radiation for two simulations, the best-performing optimized point $\{d_{\mathrm{u}}^{\mathrm{opt}}$, $d_{\mathrm{s}}^{\mathrm{opt}}$, $n_{\mathrm{p}}^{\mathrm{opt}}\}$ and the reference case without the spike are shown in Fig.~\ref{fig:parameters_betatron}. Fig.~\ref{fig:parameters_betatron}a depicts the temporal profile of the radiation and insets Fig.~\ref{fig:parameters_betatron}b and Fig.~\ref{fig:parameters_betatron}c show corresponding zoomed-in profiles around the peak values for the two studied cases. As can be seen in Fig.~\ref{fig:parameters_betatron}a, the optimized pulse arrives earlier in time compared to the reference case, and its peak value is higher by more than $25\times$. The value of $W_{50}$ increased to $W_{50}^{\mathrm{opt}}=305.0\mathrm{~nJ~sr^{-1}}$ and the value of $\tau_{50}$ dropped to $\tau_{50}^{\mathrm{opt}}=43.50~\mathrm{as}$. The optimized case also exhibits a more pronounced tail-like structure, with oscillations of a magnitude comparable to the reference pulse itself. As a consequence, the total radiated energy naturally increases, as shown in Fig.~\ref{fig:parameters_betatron}d. Importantly, not only does the low-energy part of the spectrum increase, but higher energies compared to the reference were also reached. The critical energy increases from $E_c^{\mathrm{ref}}=0.26$ keV for the reference case to $E_c^{\mathrm{opt}}=0.60$ keV for the best-performing case. 

 \subsection*{Analysis of the betatron enhancement}

\begin{figure}[h]
    \centering
 \includegraphics[width=1.05\textwidth]{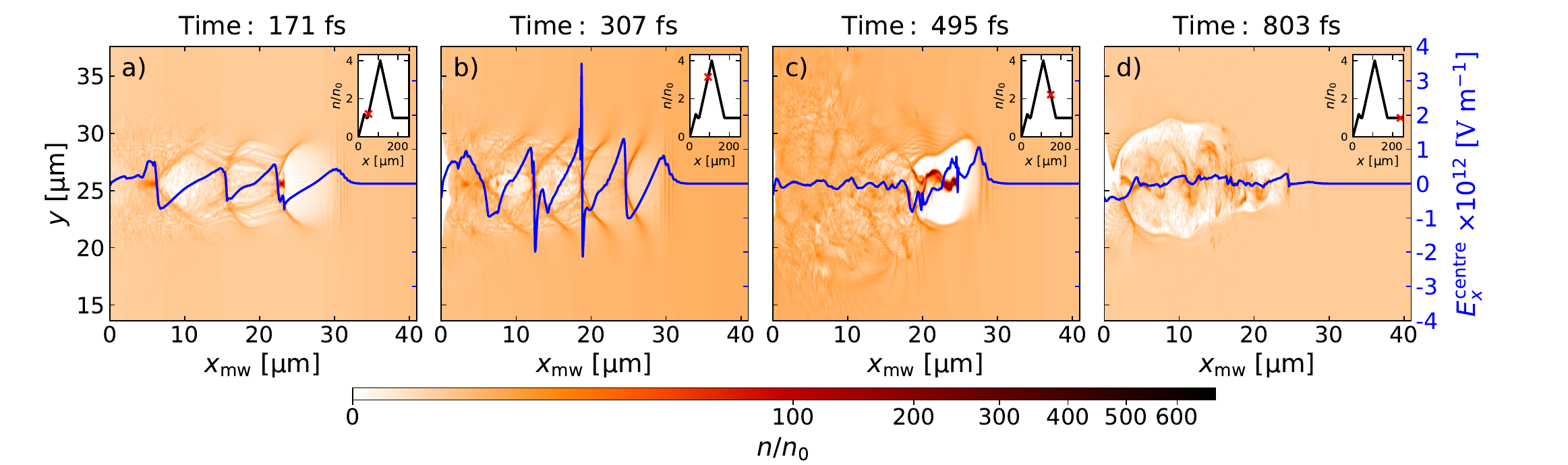}
    \caption{\textbf{Plasma density evolution for the best-performing case}. a) Early stage of propagation shortly after the first electron injection. b) Propagation through an increasing gradient of the density spike. c) A second injection is triggered while the laser pulse propagates through the decreasing gradient. d) End of the acceleration and degradation of the bubble structure. The density profiles are shown at the centre of the $z$-axis, $z=25.6~\mathrm{\upmu m}$, and $x_{\mathrm{mw}}$ is the $x$-coordinate comoving with the simulation window. For better visualization, $n/n_0$ is transformed using a power-law function with exponent $0.35$, $\left((n/n_0)/(n/n_0)_{\max}\right)^{0.35}$, where $(n/n_0)_{\max}$ corresponds to the maximum value of the normalised density $n/n_0$.
} $E_x^{\mathrm{centre}}$ (blue lines) corresponds to the longitudinal electric field at the centre of the $y$ and $z$-axis, $y=z=25.6~\mathrm{\upmu m}$.  The red crosses in the insets show the corresponding pulse position within the density gradient. The times shown above the panels are measured from when the pulse center reaches the simulation box entrance.
\label{fig:density-profile} 
\end{figure}

To obtain insight into the mechanism underlying the improvement of the betatron radiation, we analyze the dynamics of the electron beam propagating through the density spike. The evolution of the plasma density profile for the best-performing case is shown in Fig.~\ref{fig:density-profile}. Shortly after the first plasma gradient that induces the electron injection, a short electron beam is observed at the rear part of the bubble (Fig.~\ref{fig:density-profile}a). As the laser pulse propagates further through the increasing density of the spike, the bubble shrinks and electrons from the first injection are no longer sustained in the first bubble (Fig.~\ref{fig:density-profile}b). After the peak of the spike, another electron injection with an abundance of electrons is observed (Fig.~\ref{fig:density-profile}c). This injection leads to an intense attosecond betatron radiation burst with a noisy radiation tail (seen in Figs. \ref{fig:parameters_betatron}a and \ref{fig:parameters_betatron}b).
At the end of the acceleration, the original bubble structure is destroyed (Fig.~\ref{fig:density-profile}d). The leading part of the novel electron beam generates its own plasma wakefield, with a stronger electric field than the one of the laser pulse. The suppression of the laser wakefield at the centre of the $y$ and $z$ axes, $E_x^{\mathrm{centre}}$, can be seen  in Fig.~\ref{fig:density-profile}d, when compared to Fig.~\ref{fig:density-profile}a-c, where the wakefield originating from the laser pulse location is visible. The electron beam tail is accelerated in the plasma wakefield created by the leading electron beam front. A small overlap of the electron beam and laser tail was present during the final tens of femtoseconds and might have contributed to the additional radiation gain. More details on the electron and betatron beam properties, including electron energy, charge, the angular radiation spectrum, as well as the effects of multiple wakefield periods and a plasma density down-ramp at the plasma exit, are described in Supplementary Note 2.

\section*{Discussion}
Batch Bayesian optimization of attosecond betatron pulses, covering a multi-dimensional optimization space parametrizing a plasma source with a density spike, identified a parameter region with significant radiation enhancement. Comparing the performance for the batch sizes of $N=1,4,$ and $8$, we observed that batch evaluations with $N>1$ can substantially accelerate the optimization relative to the sequential approach ($N=1$), assuming that the system supports high parallelization. However, the total available resources must be carefully considered, as larger batch sizes require more simulations per iteration. In our case, $N=4$ offers the best compromise between wall-clock time, simulation count, and achieved improvement. The optimal choice of batch size, however, is inherently problem-dependent and varies depending on the computational platform and overall optimization goals. 

BBO found a region in the parameter space where the cost function increased more than 60 times compared to the reference value without the density spike. This region corresponds to a long, high-density spike positioned shortly after the initial gradient. The analysis of the PIC simulations revealed that the enhancement was due to a second, additional electron injection process induced by the density spike.

These findings highlight the efficiency of the Bayesian approach for multiparametric scans in optimization problems, especially when the number of simulations is limited or when the output is complex. In particular, revealing the efficiency of high-density down-ramp electron injection in improving the betatron pulse characteristics provides valuable guidance for experimental investigation. It should be emphasized that although the optimization converged within the explored domain, global optimality cannot be proven in principle for such multi-dimensional problems that may contain several local extrema. This limitation is inherent to such complex optimization landscapes. Choosing the batch approach instead of sequential, in particular, is well motivated if queuing and running the simulations is time consuming, or when time intensive manual processing is required. Future work could also benefit from using multi-objective Bayesian optimization techniques
to better capture trade-offs across the explored parameter space.

 \section*{Methods}\label{sec:methods}

\subsection*{Particle-in-cell simulations}

The setup for the Smilei PIC simulations was as follows:
The simulation box has dimensions $L_x \times L_y \times L_z =  40.96~\mathrm{\upmu m} \times 51.2~\mathrm{\upmu m}\times  51.2~\mathrm{\upmu m}$. A moving-window technique was applied, with a velocity of $c$ in the $x$ direction. The plasma was assumed to be fully ionized, with only electrons initialized in the simulation. The ions were considered immobile and were thus not initialized, in order to reduce computational time. The simulation grid was discretized with a resolution of $\Delta x\times \Delta y \times \Delta z = 17.8~\mathrm{nm} \times 133.3~\mathrm{nm} \times 133.3 ~\mathrm{nm}$, and the simulation timestep was $\Delta t_\text{SMILEI}=57~\mathrm{as}$. One electron macroparticle per cell was used, with regular position initialization. The Maxwell equations were solved using the Lehe scheme \cite{lehe2013numerical} in combination with a binomial filter \cite{VAY20115908}. One macroparticle represents a set of electrons with a corresponding statistical weight.
The numerical convergence tests, showing that this approach is sufficient to capture the physics for the optimization, are described in more detail in Supplementary Note 3.

\subsection*{Betatron radiation calculation}
Betatron radiation was computed using the FIKA code, which evaluates far-field emission via the Liénard–Wiechert potentials, using an approach related to previous work \cite{horny2017temporal}. FIKA requires the trajectories and momenta of the particles as input. In FIKA, only the incoherent part of the radiation is calculated, neglecting any coherence effects in the radiation. This approach is sufficient, as electrons emit radiation incoherently as betatron radiation from relativistic electron beams \cite{corde2013femtosecond}, and collective coherence effects can be neglected. Each PIC macroparticle is treated as a statistically independent radiation source, with no interference between different macroparticles. The final radiation is then obtained through a weighted summation of the radiation contributions from each PIC macroparticle \cite{pausch2018quantitatively}, with macroparticle statistical weights incorporated into the calculation.

Particle trajectories were extracted from the output of the Smilei code, sampled every $15 \Delta t_\text{SMILEI}$, where electron positions and momenta were tracked for each electron macroparticle with energy greater than 10 MeV. FIKA has an additional module that converts a Smilei output file containing particle trajectories into a format compatible with FIKA, which is also available in the FIKA GitHub repository \cite{FIKA}. The code first computed $\frac{\mathrm{d}^2W}{\mathrm{d}t \mathrm{d}\Omega}\big|_{\text{on-axis}}$, evaluated at an observer position of $(x,y,z)=(1\,\mathrm{m},0,0)$. To obtain $\frac{\mathrm{d}^2W}{\mathrm{d}\omega \mathrm{d}\Omega}\big|_{\text{on-axis}}$, the temporal resolution was refined from $15 \Delta t_\text{SMILEI}$ to $\Delta t_{\text{FIKA}} = 15 \Delta t_\text{SMILEI}/(2{{\gamma_e}_\mathrm{max}}^2)$, where ${{\gamma_e}_\mathrm{max}}$ is the maximum electron Lorentz factor along each trajectory. The radiation signal was then interpolated onto this finer temporal grid before performing the Fourier transform. 

\subsection*{Batch Bayesian Optimization}
Multi-objective Bayesian optimization is guided by an empirically defined cost function that quantifies performance across the parameter space. A Gaussian Process (GP) was used as a surrogate model to approximate the cost function. In a GP model, correlations between input points are specified by a kernel function. Here, we used the Matérn 5/2 kernel, (defined in Supplementary Note 1), which corresponds to the prior assumption that the cost function is twice mean-square differentiable. Matérn kernels enable more relaxed assumptions on the smoothness of the cost function, which allows to capture also sharper variations that might locally occur, compared to, for instance, the squared exponential (RBF) kernel, which assumes infinitely smooth functions. The acquisition function guides the search by balancing exploration and exploitation to determine where the next point should be evaluated. Here, we used the Expected Improvement (EI) acquisition function \cite{mockus1975bayesian, jones1998efficient}, which estimates the expected gain over the current best observed value. Batch points were selected using the Fantasizer approach~\cite{ginsbourger2010kriging}, which sequentially maximizes EI while updating the GP with ``fantasy" observations to avoid selecting points that are too close together. 

The optimization was implemented using the Trieste library \cite{trieste} and its Efficient Global Optimization framework.
The Bayesian optimization process was performed as follows.
(i) A batch of $N$ PIC simulations was performed with the Smilei code using selected values for \{$d_{\mathrm{u}}$, $d_{\mathrm{s}}$, $n_{\mathrm{p}}$\}. 
    Using the Smilei output of particle trajectories, $W_{50}$ and $\tau_{50}$ were obtained with the FIKA code.
    (ii) The cost function values were then computed for each point according to Eq.~\eqref{eq:cost}.
    Based on the new results, we updated the surrogate model of the cost function with GP. (iii) The acquisition function was then used to propose $N$ new points. Candidate batches were checked against the constraints in Eqs.~\eqref{eq:con1}–\eqref{eq:con4}, and if fewer than $N$ valid points were found, additional candidates were generated until a full batch of $N$ valid points was obtained.
(iv) Steps (i)–(iii) were repeated for 8 iterations.

We assumed noise-free simulation outputs from Smilei and FIKA. In the PIC Smilei simulations, macroparticles were initialized with regular positions and cold momentum, avoiding randomized initialization. We neglected potential statistical variability between runs, as all simulations were performed with the same number of particles per cell, identical resolution, and on the same computational system (16 nodes, 128 cores each), ensuring reproducible outputs, with any remaining variability expected to be negligible. The GP surrogate model was therefore trained assuming a noise-free problem, with a very small likelihood (noise) variance of $10^{-7}$ included for numerical stability.

\section*{Data availability}
The data that support the findings of this study are available from the corresponding author upon request.

\section*{Code availability}
The FIKA code developed for this work is available on GitHub under the MIT License\cite{FIKA}.

\section*{Acknowledgments}
The authors are grateful to Patrik Jansson and Ida Ekmark from Chalmers University of Technology, Miroslav Krus from Institute of Plasma Physics of the Czech Academy of Sciences, Sarah Newton from UKAEA and David Gregocki from CNR - Istituto Nazionale di Ottica for fruitful discussions. This project received funding from the Knut and Alice Wallenberg Foundation (Grants No. KAW 2020.0111 and 2023.0249). The computations were enabled by resources provided by the
National Academic Infrastructure for Supercomputing in Sweden (NAISS), partially funded by the Swedish Research Council through grant agreement No. 2022-06725, and by EuroHPC Joint Undertaking through access to Karolina at IT4Innovations (VŠB-TU), Czechia under project numberEHPC-REG-2025R01-007, together with Ministry of Education, Youth and Sports of the Czech Republic through the e-INFRA CZ (ID:90140). V.H. draws support from the European Union, the Romanian Government and the Health Program, within the project SMIS Code: 326475, and the Romanian Ministry of Research, Innovation and Digitalisation: Program Nucleu PN23210105.

\section*{Author contributions}
D.M., J.F. and I.P. conceived the main idea with inputs from V.H. and M.L. D.M., A.H. and M.L. designed the numerical simulations. D.M. and A.H. conducted the numerical simulations and analyzed the results. D.M., T.F. and I.P. wrote the manuscript with inputs from V.H. and M.L.

\section*{Competing interests}
 The authors declare no competing interests.

\section*{Supplementary Note 1: On the optimization process}

In this note, details of the batch Bayesian optimization (BBO) process are discussed. First, we show the definition of Matérn kernel used to calculate correlations between the input points. Subsequently, we show the mean values and uncertainty of the predicted model around the best-performing point found in our search.

\subsection*{Kernel definition}
Gaussian process (GP) is a probabilistic model that approximates the cost function using a mean function to predict the expected value at any point and a covariance function (kernel) to describe correlations between different points in the input space. The kernel determines the covariance matrix between all evaluated points, which is essential for making predictions in GP regression. For the batch Bayesian optimization implemented in our work, we used the Matérn kernel \cite{Matern1960}, defined as 
\begin{equation}
K(X,X') = \sigma_K^2 \, \frac{2^{1-\nu}}{\Gamma(\nu)}
\left( \sqrt{2\nu}\, r(X,X') \right)^{\nu}
K_{\nu}\!\left( \sqrt{2\nu}\, r(X,X') \right),
\end{equation}
where $K_{\nu}$ is the modified Bessel function of the second kind, $\Gamma$ denotes the gamma function, and $X=(d_{\mathrm{u}},d_{\mathrm{s}},n_{\mathrm{p}})$ and $X'=(d_{\mathrm{u}}',d_{\mathrm{s}}',n_{\mathrm{p}}')$ are points in the three-dimensional parameter space. The scaled distance $r(X,X')$ is given by
\begin{equation}
r(X,X') = \sqrt{\frac{(d_{\mathrm{u}} - d_{\mathrm{u}}')^2}{\ell_{d_{\mathrm{u}}}^2}+\frac{(d_{\mathrm{s}}-d_{\mathrm{s}}')^2}{\ell_{d_\mathrm{s}}^2}
+ \frac{(n_{\mathrm{p}} - n_{\mathrm{p}}')^2}{\ell_{n_\mathrm{p}}^2}},
\end{equation}
with $\sigma_K^2$ the variance and $\ell_{d_{\mathrm{u}}}$, $\ell_{d_{\mathrm{s}}}$, $\ell_{n_{\mathrm{p}}}$ the characteristic length scales associated with the three parameters. The smoothness of the kernel is controlled by~$\nu$. In our work, we set $\nu = 5/2$, which corresponds to a kernel whose sample functions are twice differentiable.
The kernel hyperparameters $\sigma_K^2$, $\ell_{d_{\mathrm{u}}}$, $\ell_{d_{\mathrm{s}}}$ and $\ell_{n_{\mathrm{p}}}$ were chosen by maximizing the Gaussian-process log marginal likelihood, which is a standard way to select parameters so that the surrogate model captures the data without overfitting.

\subsection*{Evaluation of the optimization model at the best-performing point}
\begin{figure*}[h!]
    \centering
   \includegraphics[width=1\textwidth]{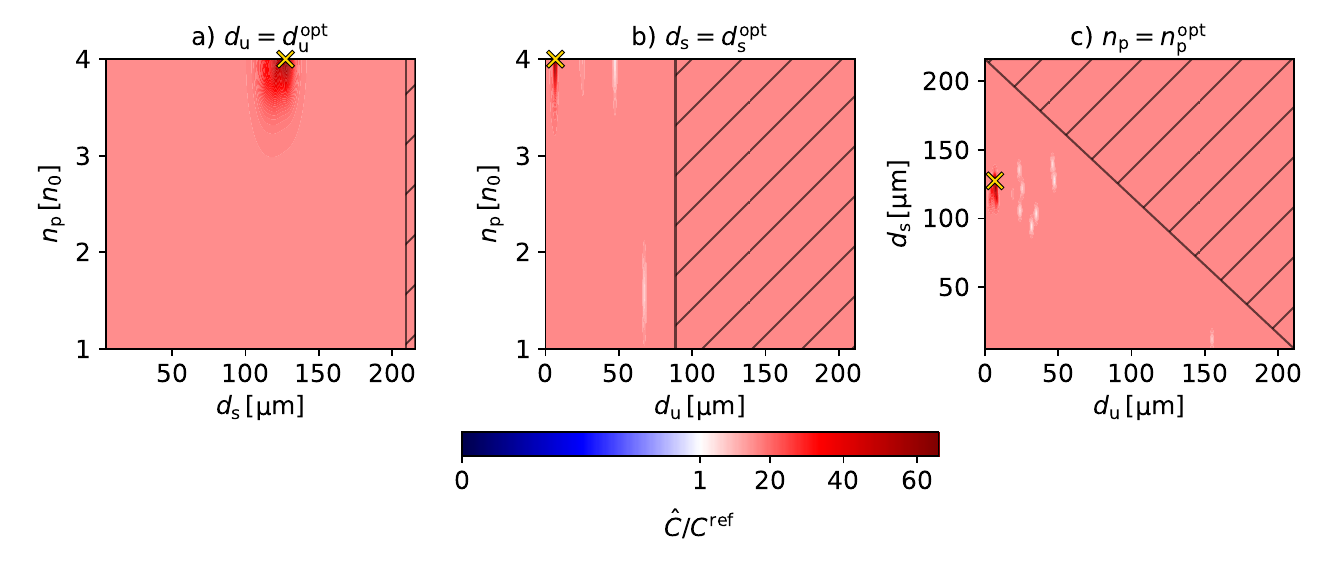}
      \includegraphics[width=1\textwidth]{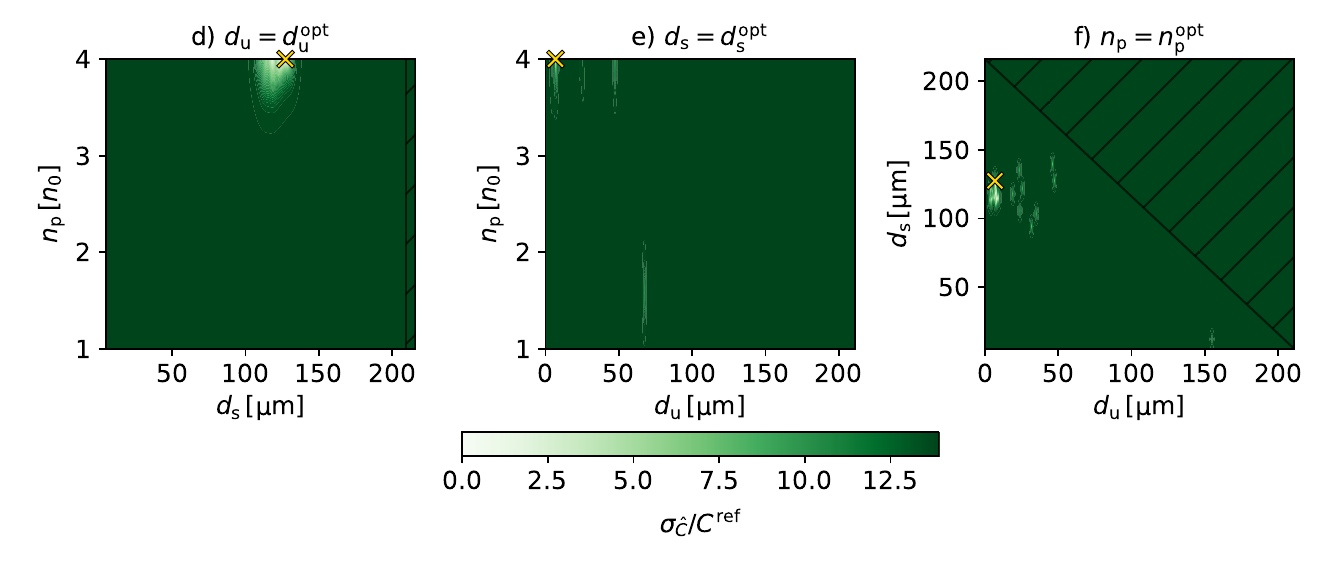}
    \caption{ 
   \textbf{ 2D maps of the surrogate model at the best-performing point.} a-c) Estimated values from the cost function model $\hat{C}$: a) 
   $d_{\mathrm{u}}^{\mathrm{opt}}$, b) $d_{\mathrm{s}}^{\mathrm{opt}}$, 
   c) $n_{\mathrm{p}}^{\mathrm{opt}}$. Red/blue shades represent an improvement/reduction of performance compared to the reference case $C^{\mathrm{ref}}$. d)-f): Values of predicted uncertainty $\sigma_{\hat{C}}$ corresponding to a)-c):  d) 
   $d_{\mathrm{u}}^{\mathrm{opt}}$, e) 
   $d_{\mathrm{s}}^{\mathrm{opt}}$, f) 
   $n_{\mathrm{p}}^{\mathrm{opt}}$. In the hashed areas, the search was forbidden due to the restriction in Eq.~(5) in the main text. The yellow cross indicates the best-performing point.} \label{fig:2Dmaps}
\end{figure*}

Here, we analyze the behaviour of the surrogate cost function $\hat{C}$ model, which estimates the performance of the cost function $C$, defined in Eq.~(1) of the main text. Figures~\ref{fig:2Dmaps}a)-c) show 2D slices of the surrogate model, which intersect each other at the best-performing point found during the whole optimization process across all the iterations and all the batches: $d_{\mathrm{u}}=d_{\mathrm{u}}^{\mathrm{opt}}=6.8~\mathrm{\upmu m}$, $d_{\mathrm{s}}=d_{\mathrm{s}}^{\mathrm{opt}}=127.4~\mathrm{\upmu m}$ and $n_{\mathrm{p}}=n_{\mathrm{p}}^{\mathrm{opt}}=4.0~n_0$. Here, $d_{\mathrm{u}}$ is  the distance between the injection of the electrons and the start of the density spike, $d_{\mathrm{s}}$ is the length of the spike, and its peak density is $n_{\mathrm{p}}$. This model is presented for the corresponding batch size where it was found, i.e. $N=8$, at the last (8th) iteration performed. The model shows strong improvement in predictions around the best-performing point. As depicted in Fig. 3 of the main manuscript, the estimates around this point predict the evaluation of $C$ very well within uncertainty. Other areas show almost uniform improvement, with $\hat{C}/C^{\mathrm{ref}}$ roughly below 20. Some small areas around certain points show performance similar to the reference case, $\hat{C}/C^{\mathrm{ref}}=1$. These points correspond to the values evaluated during the BBO exploration.

Figs.~\ref{fig:2Dmaps}d)-f) show a~corresponding uncertainty (standard deviation) of the model at the specific points. It can be seen that the model is very confident in the vicinity of the best-performing point and in narrow regions around previously evaluated points. However, the uncertainty remains large across most of the domain, indicating insufficient exploration. This behaviour was already apparent during the optimization: the search repeatedly sampled similar regions, and the cost values began to converge for all batch sizes after several iterations (see Fig.~2 of the main text). The exploration could possibly be boosted by using different acquisition and kernel functions, or possibly tuning parameters of these functions. A systematic investigation of such strategies lies beyond the scope of the present work.

\section*{Supplementary Note 2: On the properties of the electron beam and betatron radiation}

In this note, we show properties of the electron beams and the electromagnetic radiation that were not included in the main text, specifically electron energy and charge, and angular spectrum of the radiation.
\subsection*{Energy spectrum and charge evolution of the electron beam}
\begin{figure*}[htbp]
    \centering
    \includegraphics[width=0.96\textwidth]{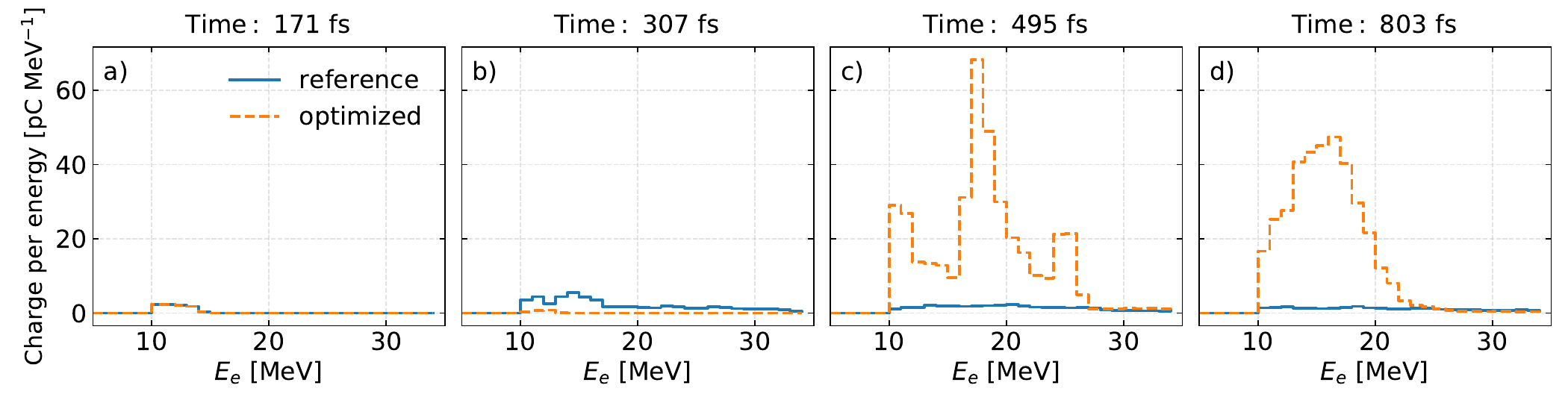}   
   \caption{Energy spectra depicted at different simulation times a) 171 fs, b) 307 fs, c) 495 fs, d) 803 fs for the reference (blue solid) and the best-performing optimized (orange dashed) cases. The times shown above the panels are measured from when the pulse centre reaches the simulation box entrance.} 
\label{fig:ene} 
\end{figure*}

To investigate how the evolution of the electron properties influences the final radiation, we present the energy spectrum at different simulation times for both the reference case and the best-performing (optimized) case, from the 7th iteration of BBO with $N=8$. The times correspond to the density evolution of the optimized case in  Fig.~5 of the main manuscript. Fig.~\ref{fig:ene} depicts the energy spectrum of electrons participating in the radiation calculation (electrons with energies $E_e \ge 10$~MeV) at different simulation times. Corresponding evolution of their total charge is summarized in Tab.~\ref{tab}. 

At 171 fs (Fig.~\ref{fig:ene}a), the spectrum corresponds to the beam from the first injection process, and the reference and optimized cases are the same. Later, at 307 fs (Fig. \ref{fig:ene}b), the charge per energy rapidly decreases in the optimized case, while it grows as the electrons are accelerating in the reference case. 
The reduction in charge is a direct consequence of the bubble contraction induced by the upward density gradient within the spike, which displaces electrons from the accelerating phase and stops the acceleration process. 

At $495$~fs (Fig.~\ref{fig:ene}c), a strong second injection occurs in the optimized case as the laser propagates through the density down-ramp. At the end of the simulation (Fig. \ref{fig:ene}d), the charge distribution is reshaped. For the optimized case, no significant additional acceleration is observed, as the main laser-driven wakefield has weakened and the wake induced by the injected electron beam dominates. The charge per energy of the reference case is significantly lower. As also shown in Tab. \ref{tab}, at the end of the simulation the final charge in the best-performing case is more than $10\times$ higher, which resulted in a rapid increase of the radiation.

\begin{table}[h!]
\centering

\begin{tabular}{l||c|c|c|c}
\hline
 & 171 fs & 307 fs & 495 fs & 803 fs \\
\hline
\hline
reference & 9 pC & 52 pC & 36 pC & 28 pC \\
optimized & 9 pC & 2 pC & 397 pC & 370 pC \\
\hline
\end{tabular}\caption{Total charge of electrons with energies $E_e\geq10$ MeV at different simulation times for the reference and optimized cases, corresponding to the spectra in Fig.~\ref{fig:ene}.}\label{tab}
\end{table}

\subsection*{Angular spectrum of the radiation}

\begin{figure*}[h]
    \centering
    \includegraphics[width=0.9\textwidth]{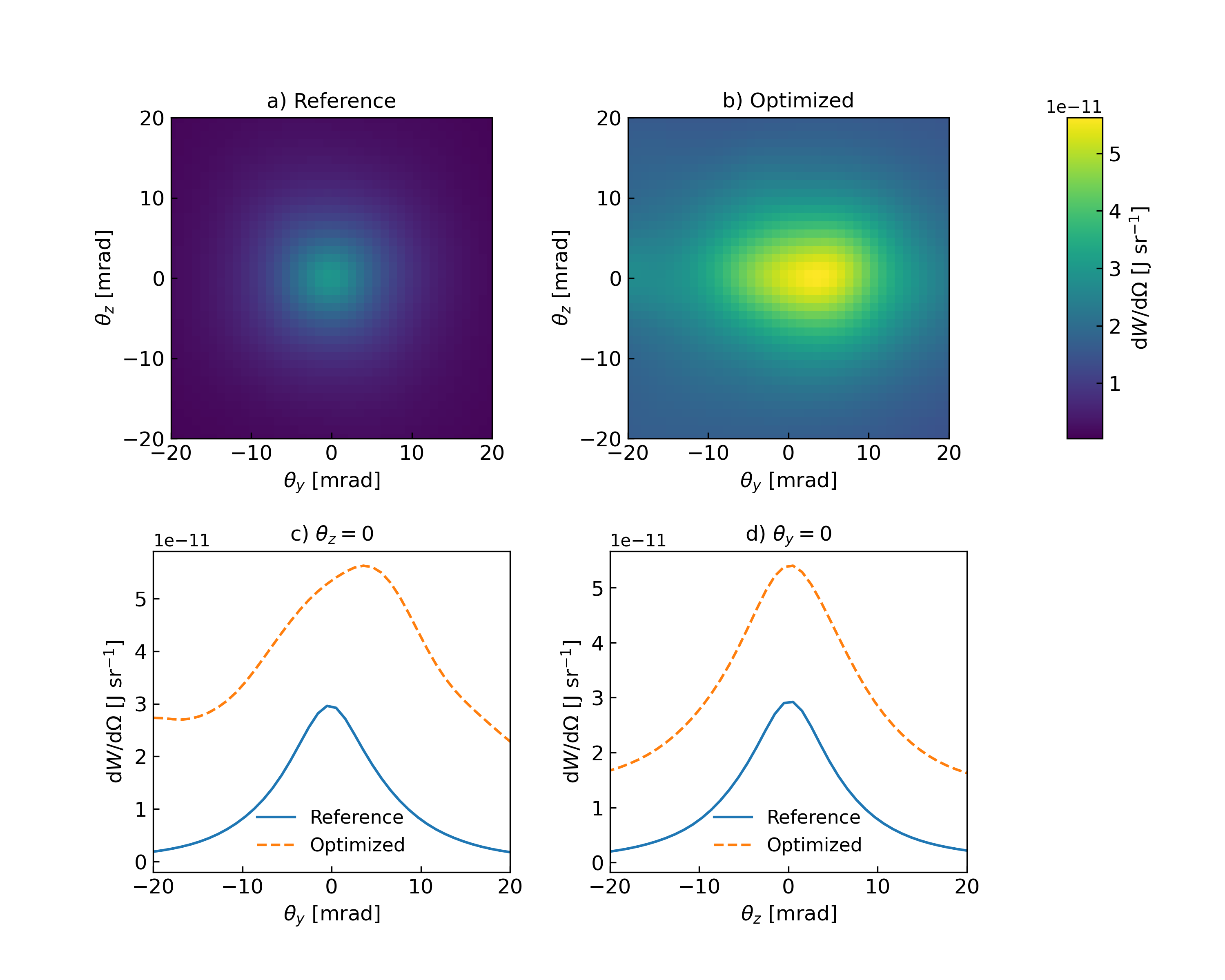}    
   \caption{\textbf{Angular spectra of the betatron radiation.} Spectra are shown for a) the reference case b) best-performing optimized case. $\theta_y$ and $\theta_z$ represent the divergences in $y$-direction and $z$-direction, respectively. $\mathrm{d}W/\mathrm{d}\Omega$ is the radiation energy $W$ per solid angle $\Omega$. On-axis lineouts of the spectra are shown at c) $\theta_z = 0$ and d) $\theta_y = 0$ for the reference case (blue solid) and the optimized case (orange dashed).}
\label{fig:ang} 
\end{figure*}

In addition to the increase in radiation energy, the angular spread of the betatron beam for the best-performing optimized case in BBO (batch size $N=8$, iteration 7) shows a wider angular spread compared to the reference, no-spike case. The root-mean-square angular spread values in both directions increased as follows: from $\sigma_{\theta_y}=10.6~\mathrm{mrad}$ to $\sigma_{\theta_y}=15.2~\mathrm{mrad}$, and from $\sigma_{\theta_z}=11.0~\mathrm{mrad}$ to $\sigma_{\theta_z}=14.4~\mathrm{mrad}$. 
The corresponding angular profiles are shown in Fig.~\ref{fig:ang}.

\subsection*{Influence of multiple wakefield periods and a density down-ramp at the plasma exit on the radiation profile}

In the main text, we focused on the first dominant betatron pulse emitted from the plasma bubble. In principle, additional electrons can be trapped in subsequent wakefield periods and contribute to the overall betatron radiation. To investigate the contribution of these later wakefield periods, we performed an additional simulation of the reference case using an extended simulation window that captures eight wakefield periods. Specifically, the longitudinal size of the simulation domain was increased from $L_x=40.96~\mathrm{\upmu m}$ to $L_x=81.92~\mathrm{\upmu m}$.

Additionally, to better approximate experimentally realistic plasma conditions, we performed an additional simulation using an extended longitudinal window that includes a plasma exit region. In contrast to the main-text simulations, where the initial plasma injection gradient is followed by a 216~$\mathrm{\upmu m}$-long uniform density plateau, the plasma profile here consists of the same injection gradient followed by a 176~$\mathrm{\upmu m}$ plateau and a subsequent 40~$\mathrm{\upmu m}$ linear down-ramp, over which the density decreases from $n_0$ to zero.

\begin{figure*}[h!]
    \centering
    \includegraphics[width=0.82\textwidth]{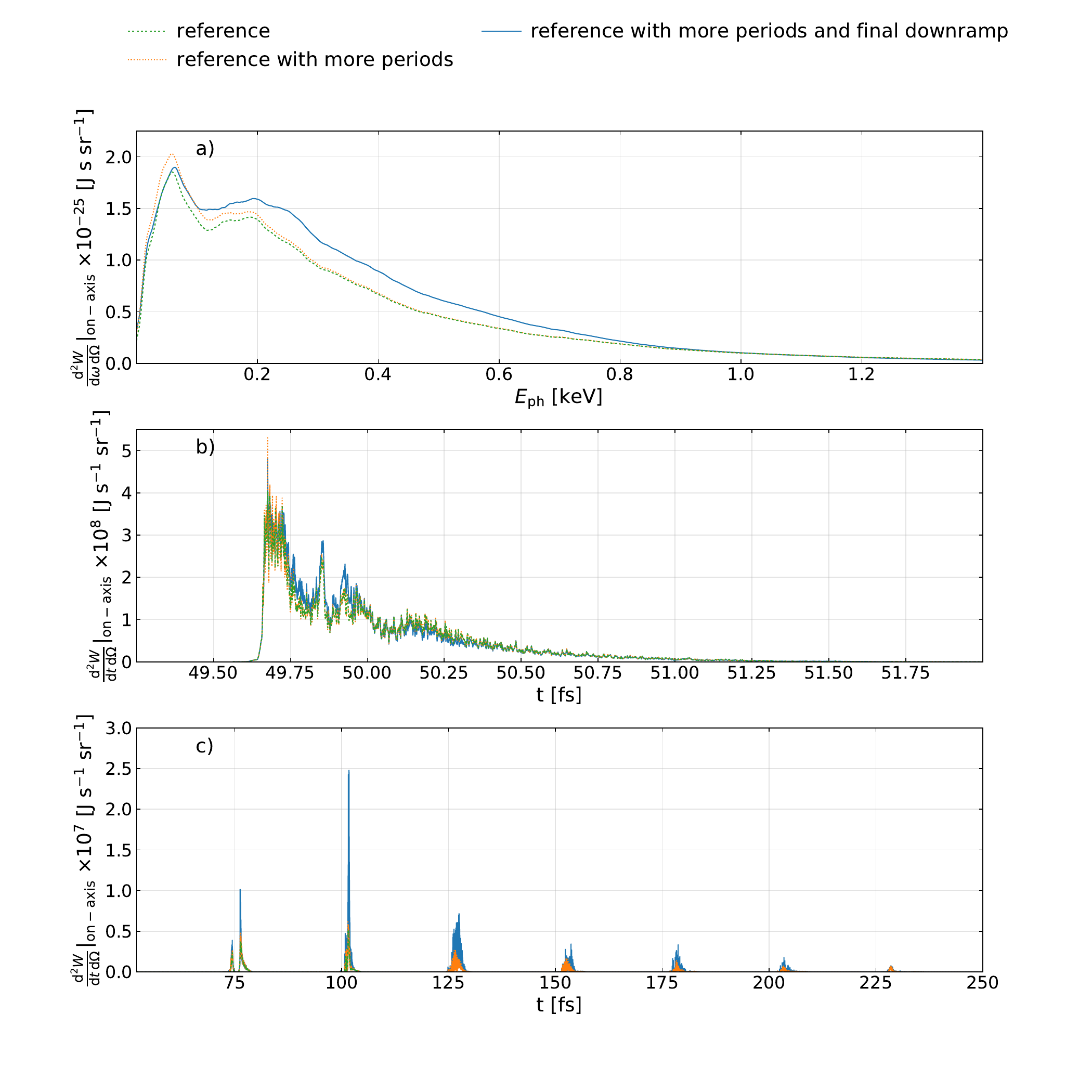}
    \vspace{-0.8cm}
   \caption{a) Dependence of the radiated energy $W$ per photon angular frequency $\omega$ per solid angle $\Omega$ on photon energy $E_{\mathrm{ph}}$. b) Temporal dependence of $W$ per observer time $t$ per $\Omega$ in the observer time window $t\in \left[ 49.25~\mathrm{fs}, 52~\mathrm{fs} \right]$. c)  Temporal dependence of $W$ per observer time $t$ per $\Omega$ in the observer time window $t\in \left[ 52~\mathrm{fs}, 250~\mathrm{fs} \right]$. The spectra are shown for the reference case without the density spike with the simulation setup from the main text (green dashed), for the reference case with the longer particle-in-cell simulation window (orange dotted), and for the reference case with the longer particle-in-cell simulation window and additional density gradient at the end of the acceleration (blue solid).}\label{fig:long_window}  
\end{figure*}

The corresponding results are shown in Fig.~\ref{fig:long_window}. As shown in Fig.~\ref{fig:long_window}c, electrons originating from each wakefield period generate distinct radiation bursts, leading to an overall increase in the energy spectrum presented in Fig.~\ref{fig:long_window}a. While extending the simulation window to include additional wakefield periods beyond eight would likely result in further radiation emission, the contribution of successive pulses is observed to decrease progressively, with the exception of the third pulse. This trend can be expected to continue for later periods. Including the density down-ramp slightly modifies the energy distribution of the emitted radiation, primarily shifting photons toward higher energies, while the overall spectral and temporal behavior remains largely unchanged.

\section*{Supplementary Note 3: On numerical convergence}
We verified the numerical convergence of Smilei particle-in-cell simulations by varying the longitudinal spatial resolution (45 versus 50 cells per laser wavelength), the transverse spatial resolution (6 versus 8 cells per laser wavelength), and the number of macroparticles per cell ($ppc=1,8,27$). Increasing the longitudinal resolution changed the resulting spectrum negligibly and the cost function increased only by 0.3\%. Increasing the transverse resolution slightly shifted the electron beam position and the timing of the betatron pulse due to a small change in the injection dynamics. However, the corresponding difference in the cost function still remained as low as 1.1\%. 

\begin{figure*}[h!]
    \centering
    \includegraphics[width=0.82\textwidth]{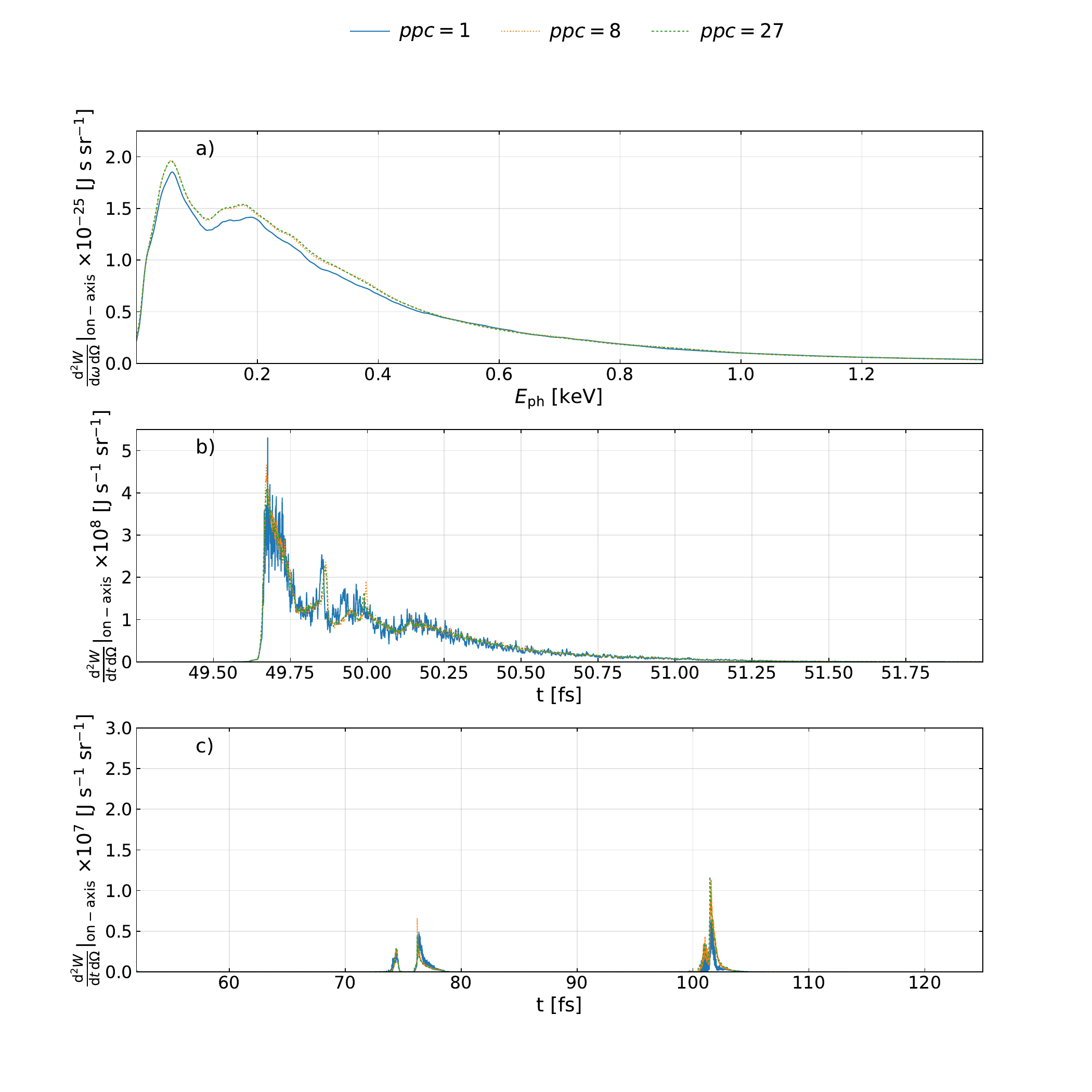}
    \vspace{-0.8cm}
   \caption{a) Dependence of the radiated energy $W$ per photon angular frequency $\omega$ per solid angle $\Omega$ on photon energy $E_{\mathrm{ph}}$. b) Temporal dependence of $W$ per observer time $t$ per $\Omega$ in the observer time window $t\in \left[ 49.25~\mathrm{fs}, 52~\mathrm{fs} \right]$. c)  Temporal dependence of $W$ per observer time $t$ per $\Omega$ in the observer time window $t\in \left[ 52~\mathrm{fs}, 125~\mathrm{fs} \right]$. The spectra are shown for the reference case without the density spike for different numbers of macroparticles per cell: a) $ppc=1$ (blue solid), b) $ppc=8$ (orange dotted), c) $ppc=27$ (green dashed).}\label{fig:macro}  
\end{figure*}

Increasing the number of macroparticles from $ppc = 1$ to $ppc = 8$ and $ppc = 27$ produced some reductions in the cost function (1.2\% and 1.7\%, respectively). The corresponding comparison of the spectra is shown in Fig. \ref{fig:macro}. Note that, in order to eliminate stochastic effects in the BBO process, we used a regular macroparticle distribution, which restricts the number of macroparticles to the third power of integers (corresponding to the three simulation dimensions). During initial BBO runs, some simulations with higher $ppc$ led to several million macroparticles being tracked (when applying an energy threshold of $E_e\geq 10~\mathrm{MeV}/c$ for tracking). The processing of radiation from these macroparticles was extremely demanding on available RAM and runtime increased dramatically (up to $17\times$ longer for Smilei and $>30\times$ longer for FIKA). 
Since the cost function integrates over the full spectrum and is robust against noise, and because all tested spatial resolutions produced nearly identical relative performance, we selected 45 cells per laser wavelength longitudinally, 6 cells per laser wavelength transversely, and $ppc = 1$ as the optimal balance between numerical stability, computational feasibility, and reproducibility in our BBO runs. 

Additional tests also confirmed minimal sensitivity to the transverse window size. Increasing the transverse window size from $51.2~\mathrm{\upmu m}$ to $64~\mathrm{\upmu m}$ produced no visible changes in the electron dynamics or radiation emission, and the cost function changed by only $0.02\%$, indicating that any numerical aperturing did not influence the results.

As an additional numerical consistency check, over the total simulation time, the total energy within the particle-in-cell simulation domain decreased by 14.1\% for the reference case without the spike, and by 39.8\% for the best-performing optimization case, consistent with energy carried out through the open boundaries. The number of electrons in the box decreased by 3.4\% for the reference case and by 71.6\% for the best-performing case, corresponding to particles leaving the computational domain throughout the interaction.

As a next step, we also varied the longitudinal-momentum threshold used to select electrons from Smilei for the radiation calculation in FIKA. It was found that thresholds of $3$ and $5~\mathrm{MeV}/c$ have cost-function differences of $\leq 0.5\%$ compared to $10~\mathrm{MeV}/c$. This is because low-energy particles contribute only marginally to the spectrum gain. However, they consume a lot of computational storage. Therefore, we picked only macroparticles with $\geq 10~\mathrm{MeV}/c$ from Smilei for the radiation calculation in FIKA.
 
For the FIKA radiation calculation, we varied the temporal step as $\Delta t_{\mathrm{FIKA}} = 25, 15,$ and $5\,\Delta t_{\mathrm{SMILEI}}$, where $\Delta t_{\mathrm{SMILEI}}$ is the Smilei timestep. All spectra were visually indistinguishable. The computational cost increased strongly for the smallest timestep: the $\Delta t_{\mathrm{FIKA}} = 5\,\Delta t_{\mathrm{SMILEI}}$ case required $233\%$ more FIKA runtime and $36\%$ more Smilei processing time compared to $\Delta t_{\mathrm{FIKA}} = 15\,\Delta t_{\mathrm{SMILEI}}$, while providing no significant improvement in the spectral output (cost function changed only by $0.6\%$). In contrast, increasing the timestep to $25\,\Delta t_{\mathrm{SMILEI}}$ reduced the computational time only slightly ($7\%$ for Smilei, $<1\%$ change for FIKA) compared to $\Delta t_{\mathrm{FIKA}} = 15\,\Delta t_{\mathrm{SMILEI}}$. The cost function decreased only by $0.4\%$. Based on these results, we selected $\Delta t_{\mathrm{FIKA}} = 15\,\Delta t_{\mathrm{SMILEI}}$ as the optimal compromise between accuracy and efficiency for BBO runs: it achieves convergence within $<1\%$ compared to $\Delta t_{\mathrm{FIKA}} = 5\,\Delta t_{\mathrm{SMILEI}}$, while avoiding the large computational load.

\begin{figure*}[t!]
    \centering
    \includegraphics[width=0.5\textwidth]{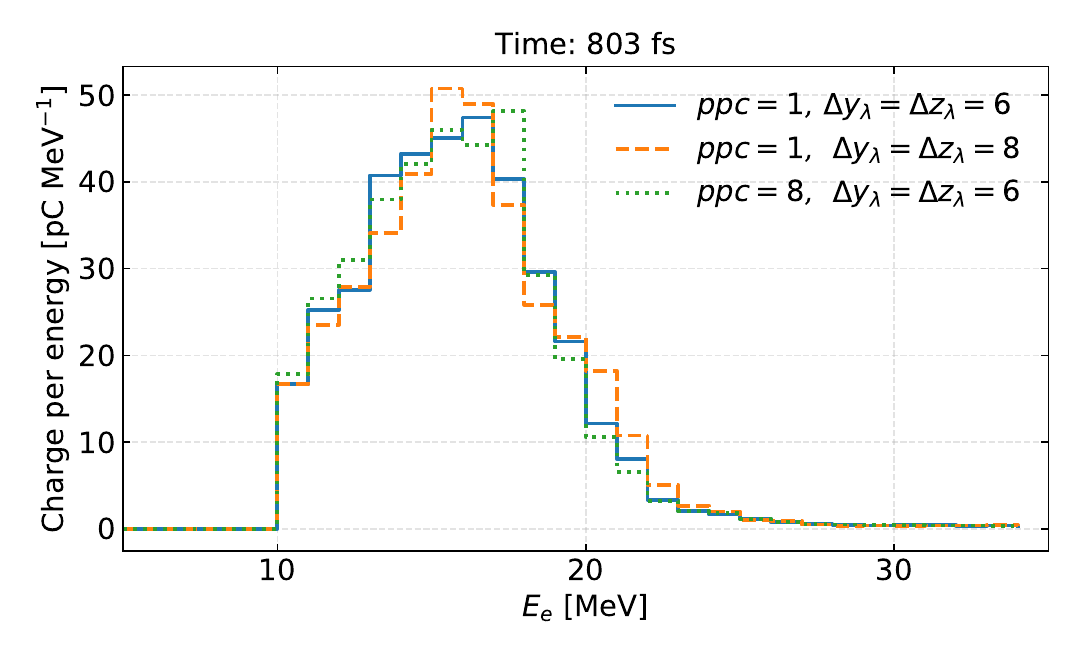}
   \caption{\textbf{Energy spectra comparison for different numerical setups}. The spectra correspond to the best-performing case at the time of 803 fs.  Following cases are depicted: $ppc = 1$ with the transverse resolution ($\Delta y_{\lambda}$ in the $y$ direction and $\Delta z_{\lambda}$ in the $z$ direction) of 6 cells per laser wavelength (blue solid), 
$ppc = 1$ with the transverse resolution of 8 cells per laser wavelength (orange dashed), and $ppc = 8$ with a transverse resolution of 6 cells per laser wavelength (green dotted).} \label{fig:energy_numerical2}
\end{figure*}

As the number of macroparticles per cell was quite low in our optimization, we additionally verified the numerical convergence of the final best-performing run. The electron energy spectrum in Fig.~\ref{fig:energy_numerical2} corresponds to a later simulation time shown in  Fig.~\ref{fig:ene}d. In one of the studied cases, we increased 
the transverse resolution to 8 cells per laser wavelength, and in the second case, we increased the number of macroparticles per cell to $ppc = 8$, while keeping the other parameters unchanged. Fig.~\ref{fig:energy_numerical2} shows that there are only slight differences in the energy distribution functions.
 The total charge of the depicted spectra is as follows: $370$ pC for the main-text case ($ppc = 1$, transverse resolution of 6 cells per laser wavelength), and $372$ pC for both other cases, $ppc = 1$ with 8 cells per laser wavelength, and $ppc = 8$ with 6 cells per laser wavelength, which indicates sufficient numerical convergence.

\bibliography{ref}
\end{document}